\documentclass{optica-article}

\journal{opticajournal} 

\articletype{Research Article}

 \usepackage{float} 
\usepackage{lineno}

\begin{document}

\title{Multipolar Decomposition of Magnetic Circular Dichroism in Arbitrarily Shaped Magneto-Dielectric Scatterers}

\author{Jhon James Hernández-Sarria,\authormark{1} João Paulo Silva Dias,\authormark{1}, Luciano Leonel Mendes,\authormark{1} Nicolò Maccaferri,\authormark{2} Osvaldo N. Oliveira Jr.,\authormark{3} and Jorge Ricardo Mejía-Salazar\authormark{1,*}}

\address{\authormark{1} National Institute of Telecommunications - Inatel, MG 37536-001 Santa Rita do Sapucaí, MG, Brazil.\\
\authormark{2} Department of Physics, Umeå University, Linnaeus väg 24, 901 87 Umeå, Sweden\\
\authormark{3} Sao Carlos Institute of Physics, University of Sao Paulo, CP 369, 13560-970 São Carlos, SP, Brazil.}

\email{\authormark{*}jrmejia@inatel.br} 


\begin{abstract*} 

Multipole expansion methods have been primarily used for analyzing the electromagnetic scattering from non-magnetic isotropic dielectric scatterers, and studies about the scattering from magnetic objects seem to be lacking. In this work, we used the multipolar expansion framework for decomposing the electromagnetic scattering by dielectric particles with magnetic properties. Magnetization current contributions were explicitly accounted for by using the vector spherical harmonics to compute the electric and magnetic multipole contributions of arbitrary order. The exact analytical expressions for the corresponding spherical multipole coefficients were employed, with the scattering efficiencies being used to distinguish the dielectric and magnetic contributions of each multipole. This enables the analysis of scattering from arbitrarily shaped, anisotropic, and inhomogeneous magnetic scatterers. It also provides a tool for studying non-reciprocal devices that exploit magnetic resonances in magnetic-dielectric materials. Calculations were made for an experimentally feasible system, namely for ferrite-based scatterers operating in the microwave regime. These materials are of interest in radio frequency (RF) applications due to their magnetic activity. We demonstrated analytically that the magnetic circular dichroism in a magnetic-dielectric scatterer in the Faraday geometry can be decomposed into individual multipole contributions. The analytical results indicate that multipole resonances associated with magnetization currents can be even stronger than multipole contributions from conventional dielectric currents. It is worth noting that these analytical results were verified through comparison with numerical results from finite element method (FEM) simulations in COMSOL Multiphysics.

\end{abstract*}

\section{Introduction}

Since Michael Faraday's observation in 1846 of the polarization rotation of an electromagnetic wave propagating parallel to a DC magnetic bias field,~\cite{Faraday1846} this so-called Faraday effect remains a subject of expanding interest.~\cite{Serber1932,Ziel1965,Cheng2019,Li2022,Niyazov2023,Merlin2024} One of the reasons for this is the non-reciprocal nature of the Faraday effect where the direction of polarization rotation depends on the relative orientation of the electromagnetic wave's propagation and the magnetic field. This non-reciprocal property has been explored in devices working across a broad spectrum of frequencies, including the visible~\cite{Xia2022,KumarJha2024,Carey2024} and microwave ranges.~\cite{Mahoney2017,Mung2019,Wang2021} Furthermore, the intrinsic chirality of the Faraday effect -- manifested as opposite rotation directions for left- (LHCP) and right-handed (RHCP) circularly polarized waves in the presence of an external magnetic field -- has been controlled in a number of applications from the microwave to the visible spectrum.~\cite{Engheta1992,Krowne1993,Krowne1993a,Yin2000,Zhou2002,Fu2003,Kuzmin2016,Iqbal2017,Wang2021a} Despite these advances, several challenges remain. One such challenge lies in the design and optimization of devices based on magnetic circular dichroism (MCD) -- the differential scattering, absorption, or extinction of light polarized to the left and right (LCP and RCP) in the presence of an external magnetic field. MCD is highly sensitive to the geometrical shape and dimensions of the magnetic scatterer, complicating the design of structures with tailored electromagnetic responses. Moreover, Faraday rotation in magnetic scatterers is often described in terms of the total effect, without quantitatively isolating the contributions of individual photonic resonances. This lack of decomposition hinders both physical insight and practical exploitation of these phenomena in advanced applications.\\

The emergence of three-dimensional (3D) nanomagnetism has brought the need for analytical formulations that can accurately account for multipolar contributions arising from magnetization currents.\cite{Gubbiotti_2025} To address this, magnetization effects have been integrated into far-field scattering calculations through a full-wave electromagnetic (EM) theory. The latter framework enables the multipole decomposition of two-dimensional (2D) anisotropic, inhomogeneous, and arbitrarily shaped cylindrical structures.\cite{Loulas_2025} The method involves expanding the scattered electric field into cylindrical divergence-free vector wave functions (CVWFs), with the expansion coefficients linked to the multipolar response. However, this approach remains constrained to structures with cylindrical symmetry. Other efforts have employed multipole decomposition techniques based on spherical harmonics expansion to model radiated fields.\cite{PhysRevB.100.125415} The exact multipole moments associated with wave scattering by magnetic particles have been explicitly derived, with closed-form expressions available up to the quadrupole order.\cite{Evlyukhin_2023}. Although Ref. \cite{PhysRevB.100.125415} introduces a methodology for deriving exact higher-order multipoles that explicitly incorporate magnetization current contributions, analytical expressions have so far been reported only up to the hexadecapole order. Notably, a complete tensorial representation of the magnetic hexadecapole moment remains unavailable—only a symmetrized primitive form has been presented in the literature. Deriving the full tensor is challenging due to its inherent mathematical complexity. While these advanced methods have deepened the theoretical foundation of multipolar analysis, they are often limited by the rapidly growing complexity of higher-order multipole terms.\cite{Zenin_2020}\\

In contrast to the frameworks described above, the conventional multipole theory based on vector spherical harmonics, originally developed in Gaussian units,\cite{Hansen_1935, Heitler_1936, KRAMERS1943261, Casimir_1954, Devaney_1974} and later adapted to the International System of Units (SI) in Jackson’s Classical Electrodynamics,\cite{jackson_classical_1999} accommodates arbitrary 3D geometries and remains analytically tractable even for higher-order multipoles. Building on this framework, Grahn et al.\cite{Grahn2012} derived expressions widely used in photonics\cite{Yang_2018,Review_Alaee_2019, James_ACS_2021, Dezert_2019, james_PRL_2021} to compute electric and magnetic spherical multipole coefficients for non-magnetic, isotropic, and linear media. In this work, we used the exact expressions for the electric and magnetic spherical multipole coefficients of wave scattering by magnetic particles to numerically investigate the scattered electromagnetic waves by magnetized objects having both relative permittivity ($\hat \varepsilon$) and permeability ($\hat \mu$) described by tensors forms. We disentangled the contributions of electric and magnetic photonic resonances to the magnetic circular dichroism (MCD) in magnetized dielectric scatterers, thus providing a versatile framework for obtaining deeper insights into the far-field behavior of dielectric–magnetic scatterers. Departing from the traditional multipole expansion,\cite{Grahn2012} we account numerically for the far-field effects of magnetization-induced bound currents—a contribution that becomes particularly relevant in the microwave regime, where ferrimagnetic materials exhibit anisotropic permeability tensors under a static magnetic bias.\cite{pozar_2012} By computing  the individual multipole components associated with both polarization and magnetization currents, we demonstrate that their sum converges to results obtained from full-wave electromagnetic simulations, such as those performed in COMSOL Multiphysics. 

\section{Theoretical Framework}

The scattering properties are analyzed for a scatterer consisting of an anisotropic and non-homogeneous linear material of arbitrary shape and size. The system is characterized by tensors for relative permittivity, $ \hat{\varepsilon}_p $, and permeability, $ \hat{\mu}_p $. The scattering system is embedded in a surrounding medium that is uniform, isotropic, and lossless, described by constant scalar values for relative permittivity, $ \varepsilon_s $, and permeability, $ \mu_s $. To analyze the electromagnetic behavior of this system systematically, we employ Maxwell's equations for time-harmonic fields,
\begin{eqnarray} 
&& \nabla \cdot \mathbf{D} \left( \mathbf{r} \right)  = 0 ,  \\
&& \nabla \cdot \mathbf{B} \left( \mathbf{r} \right)  = 0 , \\
&&\nabla \times \mathbf{E} \left( \mathbf{r} \right) = i \omega  \mathbf{B} \left( \mathbf{r} \right) \; , \\
&& \nabla \times \mathbf{H} \left( \mathbf{r} \right) = - i \omega \mathbf{D} \left( \mathbf{r} \right) \; ,
\end{eqnarray}
with complex electric and magnetic fields $ \mathbf{E} e^{-i \omega t } $ and $ \mathbf{H}  e^{-i \omega t } $, respectively. Here $\omega$ is the angular frequency, $t$ represents time. The dielectric displacement and magnetic induction are given by the following constitutive relations $ \mathbf{D} = \varepsilon_0 \hat{\varepsilon} \left( \mathbf{r} \right) \mathbf{E} $ and $ \mathbf{B} = \mu_0 \hat{\mu} \left( \mathbf{r} \right) \mathbf{H} $ (\textbf{r} is the position vector), respectively.  The constants $ \varepsilon_0 $ and $ \mu_0 $ correspond to the vacuum permittivity and permeability. \\

Following the approach outlined in Ref. \cite{PhysRevB.100.125415}, Maxwell's equations can be formulated for the entire spatial domain using the piecewise tensors
\begin{equation}
\Delta \hat{\varepsilon} = \left\{ \begin{array}{lc} \hat{\varepsilon}_p  - \varepsilon_s  \mathbb{I}  &  \in V_p, \\ \\ 0 & \notin V_p, \end{array} \right .
\end{equation}
\begin{equation}
\Delta \hat{\mu} = \left\{ \begin{array}{lc} \hat{\mu}_p  - \mu_s  \mathbb{I}  &  \in V_p, \\ \\ 0 & \notin V_p. \end{array} \right .
\end{equation}
Here, $ V_p $ is the volume of the scatterer, and $ \mathbb{I} $ denotes the $ 3 \times 3 $ unit tensor. Using these definitions, Maxwell's equations take the following form
\begin{eqnarray}
\nabla \times \mathbf{E} & = & i\omega \mu_0\left(\mu_s\mathbb{I}+\Delta \hat{\mu}\right)\mathbf{H}\; , \\
\nabla \times \mathbf{H} & = & -i\omega \varepsilon_0 \left(\varepsilon_s\mathbb{I}+\Delta \hat{\varepsilon}\right)\mathbf{E},
\end{eqnarray}
and the wave equations for the electric and magnetic fields along the entire space become\cite{Evlyukhin_2023}
\begin{eqnarray}
\label{wave_electric}
\nabla \times \left( \nabla \times \mathbf{E} \right) - k^2 \mathbf{E} &=& \omega^2 \mu_0 \mu_s \mathbf{P} + i \omega \mu_0 \nabla \times \mathbf{M}, \\
\label{wave_magnetic}
\nabla \times \left( \nabla \times \mathbf{H} \right) - k^2 \mathbf{H} &=& \omega^2 \mu_0 \varepsilon_0 \varepsilon_s  \mathbf{M} - i \omega \nabla \times  \mathbf{P},
\end{eqnarray}
with $ k = \omega \sqrt{\mu_0 \varepsilon_0 \mu_s \varepsilon_s} $ for the wave number of the surrounding dielectric medium, and the electric ($ \mathbf{P} $) and magnetic ($ \mathbf{M} $) vector polarization densities
\begin{equation}
\mathbf{P} = \varepsilon_0 \left( \hat{\varepsilon}_p  - \varepsilon_s  \mathbb{I} \right) \mathbf{E} \; , ~~~~~~  \mathbf{M} = \left( \hat{\mu}_p  - \mu_s  \mathbb{I} \right) \mathbf{H}. 
\end{equation}

Using the Lagrange's formula $ \nabla \times \left( \nabla \times \mathbf{A} \right) = \nabla \left( \nabla \cdot \mathbf{A} \right) - \nabla^2 \mathbf{A} $, which holds for any well-behaved vector field $ \mathbf{A}$, the corresponding inhomogeneous Helmholtz equations are obtained as
\begin{equation}
\left( \nabla^2 + k^2  \right) \mathbf{E} = - \omega^2 \mu_0 \mu_s \mathbf{P} - i \omega \mu_0 \nabla \times \mathbf{M} - \frac{1}{\varepsilon_0 \varepsilon_s} \nabla \left( \nabla  \cdot  \mathbf{P} \right),
\end{equation}
\begin{equation}
\left( \nabla^2  + k^2  \right) \mathbf{H} = - \omega^2 \mu_0 \varepsilon_0 \varepsilon_s  \mathbf{M} + i \omega \nabla \times  \mathbf{P} - \frac{1}{\mu_s} \nabla \left( \nabla  \cdot  \mathbf{M} \right), 
\end{equation}

The nonzero divergence of the electric and magnetic fields gives rise to the extra terms that depend on the gradients of the divergence of the magnetization and polarization densities. To compute the spherical multipole coefficients, the radial components of the electromagnetic fields must be determined,\cite{Casimir_1954} which can be achieved by taking the scalar product of the derived vector differential equations with the position vector $ \mathbf{r} $. Utilizing the mathematical identities $ \mathbf{r} \cdot \nabla^2 \mathbf{A} = \nabla^2 \left( \mathbf{r} \cdot \mathbf{A} \right) - 2 \nabla \cdot \mathbf{A}  $  and $ \mathbf{r} \cdot \left( \nabla \times \mathbf{A} \right) = - \nabla \cdot \left( \mathbf{r} \times \mathbf{A} \right) $ the inhomogeneous Helmholtz equations governing the radial components of the electromagnetic fields are derived as follows,
\begin{eqnarray}
\label{electric_radial}
 \left( \nabla^2 + k^2  \right) \left( \mathbf{r}  \cdot  \mathbf{E} \right) & \hspace{-1mm} = \hspace{-1mm} & i \omega \mu_0 \nabla \cdot \left( \mathbf{r}  \times \mathbf{M} \right) - \omega^2 \mu_0 \mu_s \mathbf{r}  \cdot \mathbf{P} - \frac{1}{\varepsilon_0 \varepsilon_s} \left(  2 + r \frac{\partial}{\partial r} \right)  \left( \nabla \cdot \mathbf{P} \right) \; , \\ \label{magnetic_radial}
\left( \nabla^2 + k^2  \right) \left( \mathbf{r}  \cdot  \mathbf{H} \right) & \hspace{-1mm} = \hspace{-1mm} & - i \omega \nabla \cdot \left( \mathbf{r} \times  \mathbf{P} \right) - \omega^2 \mu_0 \varepsilon_0 \varepsilon_s \mathbf{r}  \cdot \mathbf{M} - \frac{1}{\mu_s}  \left( 2 + r \frac{\partial}{\partial r} \right)\left( \nabla \cdot \mathbf{M} \right) \; . 
\end{eqnarray}
These differential equations define the radial dependence of the electric and magnetic fields in the presence of $ \mathbf{P} $ and $ \mathbf{M} $. It is worth emphasizing that these differential equations take into account all possible bound current contributions (i.e. dielectric and magnetic) within an arbitrarily shaped, anisotropic, and inhomogeneous scatterer into the conventional multipole theory. Note that the total electromagnetic fields outside the particle can be expressed as a superposition of the scattered and incident fields, i.e.,
\( \mathbf{E} = \mathbf{E}_s + \mathbf{E}_{\text{inc}} \), and \(\mathbf{H} = \mathbf{H}_s + \mathbf{H}_{\text{inc}}. \) As it is well-known, wave fields propagating in a lossless and source-free medium satisfy the homogeneous Helmholtz equation, consequently, the scattered fields must also obey the same differential equations ---specifically, equations~\eqref{electric_radial}–\eqref{magnetic_radial}--- since the incident fields themselves are homogeneous solutions of the Helmholtz equation. The solutions to Eqs.~\eqref{electric_radial}–\eqref{magnetic_radial} for the scattered fields can be obtained directly using the Green's function formalism, which provides a powerful and systematic framework for solving inhomogeneous differential equations. This method enables accurate characterization of electromagnetic field behavior in complex material systems, thereby broadening its applicability to a wide range of physical scenarios.

\section{Spherical multipole coefficients}

The scattered electromagnetic field by a scattering system (which is surrounded by a homogeneous, isotropic and lossless medium) that is interacting with a plane wave can be written in spherical coordinates as an multipole expansion as,\cite{Grahn2012}
\begin{eqnarray}
\nonumber
\mathbf{E}_{s} \left( r, \theta, \varphi \right) & \hspace{-1mm} = \hspace{-1mm} & E_0 \sum_{l = 1}^{\infty} \sum_{m = - l}^{l}  i^l \left[ \pi \left( 2 l + 1 \right) \right]^{1/2} 
 \times \bigg\{ \frac{1}{k} a_{E} \left( l, m \right) \nabla \times \left[ h^{\left( 1 \right)}_l \left( kr \right)  \mathbf{X}_{lm} \left( \theta , \varphi  \right)  \right]  \\
 && +  a_{H} \left( l, m  \right) h^{\left( 1 \right)}_{l} \left( kr \right) \mathbf{X}_{l,m} \left( \theta, \varphi \right)  \bigg\} \; ,  
\end{eqnarray}

\begin{eqnarray}
\nonumber
\mathbf{H}_{s} \left( r, \theta, \varphi \right) & \hspace{-1mm} = \hspace{-1mm} & \frac{E_0}{ \eta } \sum_{l = 1}^{\infty} \sum_{m = - l}^{l}  i^{l-1} \left[ \pi \left( 2 l + 1 \right) \right]^{1/2} 
\times \bigg\{ \frac{1}{k} a_{H} \left( l, m \right) \nabla \times \left[ h^{\left( 1 \right)}_l \left( kr \right)  \mathbf{X}_{lm} \left( \theta , \varphi \right) \right] \\
&& + a_{E} \left( l, m \right) h^{\left( 1 \right)}_{l} \left( kr \right) \mathbf{X}_{l,m} \left( \theta, \varphi \right)  \bigg\} \; , 
\end{eqnarray}
where \( \mathbf{X}_{lm} \left( \theta , \varphi \right) \) represents the normalized vector spherical harmonics (VSH), which can be constructed using the spherical harmonic functions $  Y_{lm} \left( \theta, \phi \right) $ as follows,\cite{lambert_1978}
\begin{equation}
 \mathbf{X}_{lm} = \left[ \frac{i}{ \sqrt{ l \left( l + 1 \right) } \sin \theta} \frac{\partial Y_{lm} }{\partial \varphi} \right] \hat{\theta} + \left[ \frac{- i  }{ \sqrt{l \left( l + 1 \right) } } \frac{\partial Y_{lm}}{\partial \theta} \right]  \hat{\varphi} \; .
\end{equation} 

Where, \( \hat{\theta} \) and \( \hat{\varphi} \) denote the unit vectors in the polar and azimuthal directions, respectively, in the spherical coordinate system. \( h^{\left( 1 \right)}_{l} \left( kr \right) \) denotes the spherical Hankel functions of the first kind. The wavenumber \( k \) corresponds to that of the surrounding dielectric medium, which has an impedance \( \eta = \sqrt{\mu_s / \varepsilon_s} \). The normalizing constants of the multipole fields are chosen so that the expressions for the scattering and extinction cross sections can be written in a compact way. The electric and magnetic spherical multipole coefficients $a_E(l, m)$ and $a_H(l, m)$ represent the strength of each vector spherical harmonic in the multipole expansion. Using the orthogonal relations that satisfy the vector spherical harmonics,\cite{jackson_classical_1999, lambert_1978} the spherical multipole coefficients can be calculated using the radial component of the scattered EM fiels as,\cite{Grahn2012}
\begin{eqnarray}
a_E \left( l , m \right) = - \frac{\left( - i \right)^{l - 1} k  }{ h^{1} \left( kr \right) E_0  \left[ \left( 2 l + 1 \right)  l \left( l + 1 \right) \right]^{1/2}}  
\times \int_{0}^{\pi} \int_{0}^{2 \pi} Y_{lm}^{*} \left( \theta, \varphi \right) \; \mathbf{r} \cdot \mathbf{E}_s \; d \Omega \; ,
\end{eqnarray}
\begin{eqnarray}
 a_H \left( l , m \right) = \frac{\left( - i \right)^{l} \eta k }{ h^{1} \left( kr \right) E_0  \left[ \left( 2 l + 1 \right)  l \left( l + 1 \right) \right]^{1/2}}  \times \int_{0}^{\pi} \int_{0}^{2 \pi} Y_{lm}^{*} \left( \theta, \varphi \right) \; \mathbf{r} \cdot \mathbf{H}_s \; d \Omega \; ,
 \label{multipolo_coeficiente_magnetico}
\end{eqnarray}
where the radial components of the electromagnetic fields are derived using the Green function formalism (see Eqs.~(\ref{electric_radial}) and (\ref{magnetic_radial})) with the boundary condition of outgoing waves at infinity. Moreover, using the spherical wave projection of the Green's functions,\cite{jackson_classical_1999}
\begin{equation}
\frac{1}{4 \pi} \int_{\Omega} Y_{l,m} \left( \theta, \varphi \right)  \frac{e^{i k \left| \mathbf{r} - \mathbf{r}^{\prime} \right|}}{ \left| \mathbf{r} - \mathbf{r}^{\prime} \right|} d \Omega = i k h^{(1)}_l \left( k r \right) j_l \left( k r^{\prime} \right) Y_{lm}^{*} \left( \theta^{\prime} , \varphi^{\prime} \right) \; ,
\end{equation}

It is worth noting that the spherical wave expansion of the Green's function used in this decomposition technique is valid only when the observation point $ \mathbf{r} $, at which the electromagnetic fields are evaluated, lies outside a spherical surface that completely encloses the scattering system.~\cite{jackson_classical_1999} Under this condition, the electric and magnetic spherical multipole coefficients for a dielectric–magnetic scatterer embedded in a lossless dielectric medium reduce to
\begin{eqnarray}
\hspace{-5mm} a_E \left( l , m \right) & \hspace{-1mm} = \hspace{-1mm} & \frac{ - \left( - i \right)^{l} k^2}{E_0  \left[ \left( 2 l + 1 \right)  l \left( l + 1 \right) \right]^{1/2}} \int_{V} j_l \left( k r \right) Y_{lm}^{*} \left( \theta , \varphi \right) \\
&& \bigg\{ i \omega \mu_0 \nabla \cdot \left( \mathbf{r} \times \mathbf{M} \right) - \omega^2 \mu_0 \mu_s \mathbf{r} \cdot \mathbf{P}  - \frac{1}{ \varepsilon_0 \varepsilon_s } \left( 2 + r \frac{\partial }{\partial r} \right) \nabla \cdot \mathbf{P} \bigg\} d^{3} r,  
\end{eqnarray}

\begin{eqnarray}
\hspace{-5mm} a_H \left( l , m \right) & \hspace{-1mm} = \hspace{-1mm} & \frac{  \left(-i \right)^{l + 1} \eta k^2}{E_0  \left[ \left( 2 l + 1 \right)  l \left( l + 1 \right) \right]^{1/2}}  \int_{V} j_l \left( k r \right) Y_{l,m}^{*} \left( \theta , \varphi \right) \\
&&\bigg\{ - i \omega \nabla \cdot \left( \mathbf{r} \times \mathbf{P} \right) - \omega^2 \mu_0 \varepsilon_0 \varepsilon_s \mathbf{r} \cdot \mathbf{M} - \frac{1}{ \mu_s }  \left( 2 + r \frac{\partial }{\partial r} \right) \nabla \cdot \mathbf{M}  \bigg\} d^{3} r.
\end{eqnarray}

These expressions quantify the strengths of the various multipolar components of the scattered field outside the particle in terms of volume integrals involving the polarization and magnetization densities. In fact, outside of a spherical surface that fully encloses the scatterer. In particular, these equations permit the analysis of the influence from magnetization currents on the far-field properties of the system. In ferrimagnetic materials, such currents arise from spin precession effects. Hence, the radiation scattered due to spin precession in ferrites can be decomposed into its multipole contributions. In the following relationships, we present reformulated expressions optimized for numerical implementation. These adaptations eliminate the need for numerical differentiation of the induced polarization and magnetization densities, while clearly separating the contributions from the two types of bound current densities ---namely, polarization and magnetization currents. By expressing their respective contributions to the spherical multipole coefficients independently, we can pinpoint the specific physical mechanisms responsible for the distinct features observed in the far-field properties.
\begin{eqnarray}
&& a_{E} \left( l , m \right) =  a^{\text{P}}_{E} \left( l , m \right)
+ a^{\text{M}}_{E} \left( l , m \right) \; , \\
&& a_{H} \left( l , m \right) =  a^{\text{P}}_{H} \left( l , m \right)
+ a^{\text{M}}_{H} \left( l , m \right) \; .
\end{eqnarray}

The spherical multipole coefficients corresponding to the fields scattered by particle are expressed as

\begin{eqnarray}
\nonumber
a^{\text{P}}_{E} \left( l , m \right) && = \frac{ \left( - i \right)^{l - 1} k^2 \eta O_{lm}}{ E_0 \left[ \pi \left( 2 l + 1 \right) \right]^{1/2} } \int e^{-im\varphi} \bigg\{ \left[ \Psi_l \left( k r \right)  + \Psi_l^{\prime \prime} \left( k r \right) \right] P^{m}_{l} \left( \cos \theta \right) \hat{r} \cdot \mathbf{J}_{\text{P}} \\
&& + \frac{\Psi^{\prime} \left( k r \right)}{k r} \left[ \tau_{lm} \left( \theta \right) \hat{\theta} \cdot \mathbf{J}_{\text{P}} - i \pi_{lm} \left( \theta \right)  \hat{\varphi} \cdot \mathbf{J}_{\text{P}} \right] \bigg\} ~ d^3 r \; , \\
a^{\text{M}}_{E} \left( l , m \right) && =  - \frac{ \left( - i \right)^{l - 1} k^3 \eta O_{lm}}{ \mu_s E_0 \left[ \pi \left( 2 l + 1 \right) \right]^{1/2} } \int e^{-i m \varphi} j_{l} \left( k r \right) \bigg\{ i \pi_{lm} \left( \theta \right) \hat{\theta} \cdot \mathbf{M} + \tau_{lm} \left( \theta \right) \hat{\varphi} \cdot \mathbf{M} \bigg\} ~ d^3 r \; , \\ 
a^{\text{P}}_{H} \left( l , m \right) &&= \frac{ \left( - i \right)^{l + 1} k^2 \eta O_{lm}}{ E_0 \left[ \pi \left( 2 l + 1 \right) \right]^{1/2} } \int e^{-im\varphi} j_{l} \left( k r \right) \bigg\{ i \pi_{lm} \left( \theta \right) \hat{\theta} \cdot \mathbf{J}_{\text{P}} + \tau_{lm} \left( \theta \right) \hat{\varphi} \cdot \mathbf{J}_{\text{P}} \bigg\} ~ d^3 r \; .\\ \nonumber
 a^{M}_{H} \left( l , m \right) &&= - \frac{ \left( - i \right)^{l + 1} k^3 \eta O_{lm}}{ \mu_s E_0 \left[ \pi \left( 2 l + 1 \right) \right]^{1/2} } \int e^{-im\varphi} \bigg\{ \left[ \Psi_l \left( k r \right)  +  \Psi_l^{\prime \prime} \left( k r \right) \right] P^{m}_{l} \left( \cos \theta \right) \hat{r} \cdot \mathbf{M} \\
&& + \frac{\Psi^{\prime} \left( k r \right)}{k r} \left[ \tau_{lm} \left( \theta \right) \hat{\theta} \cdot \mathbf{M} - i \pi_{lm} \left( \theta \right)  \hat{\varphi} \cdot \mathbf{M} \right] \bigg\} ~ d^3 r \; .
\end{eqnarray}

Integration by parts has been employed to derive these expressions (for details, see Appendix), as directly computing the spatial derivatives of the polarization and magnetization densities can be numerically cumbersome. Additionally, the induced polarization current density, $ \mathbf{J}_{\text{P}} = - i \omega \mathbf{P} $ has been introduced to express the spherical multipole coefficients in a more convenient form. These expressions allow us to identify the multipolar contribution of the Cartesian toroidal multipole within the long-wavelength approximation (LWA).\cite{PhysRevB_Tiago_2024} Furthermore, to streamline the previous expressions, the following parameters and functions have been defined
\begin{eqnarray}
&&  \Xi_{lm} =  \sqrt{\frac{2l+1}{4\pi}\frac{\left(l-m\right)!}{\left(l+m\right)!}} ~ , \\
&&  O_{lm} = \frac{\Xi_{lm}}{ \sqrt{ l \left( l + 1 \right) }} ~ , \\
&&  \pi_{lm} \left( \theta \right)  =  \frac{m}{\sin \theta} P^{m}_{l} \left( \cos \theta \right) ~ , \\
&&  \tau_{lm} \left( \theta \right)  = \frac{d}{d \theta} P^{m}_{l} \left( \cos \theta  \right) ~  .
\end{eqnarray}

The parameters $ \Xi_{lm} $ and $ O_{lm} $ represent the normalization constants of the spherical harmonics and vector spherical harmonics, respectively. In addition, the Riccati-Bessel function $ \Psi_{l} \left( k r \right) = k r j_l  \left( k r \right) $ has been used.

\section{Cross section calculations}

The scattering and extinction cross sections associated with each polarization density and magnetization density are calculated independently to evaluate the impact of each bound current on the far-field properties. From scattering theory, the total scattered power $ W_{sca} $ can be determined using the angular components of the scattered electromagnetic fields, written as,\cite{bohren1998}
\begin{eqnarray}
W_{sca} = \frac{1}{2} \Re \bigg\{ \int_{\omega}  \left[ E_{s \theta} H^{*}_{s \varphi} -  E_{s \varphi} H^{*}_{s \theta} \right] r^2 \; d \Omega\bigg\} .
\end{eqnarray}

Following a standard approach, the scattering cross section can be decomposed into its individual contributions --namely, the fields scattered by polarization and magnetization current densities.~\cite{Evlyukhin_2023} Consequently, the total scattering cross section is 
\begin{equation}
\label{totalscaPM}
\sigma_{sca} = \sigma_{sca}^{\text{P}} + \sigma_{sca}^{\text{M}} + 2 \Re\{ \sigma_{sca}^{\text{PM}}\} .
\end{equation}
where $\sigma_{sca}^{P}$ and $\sigma_{sca}^{M}$ are the contributions from the dielectric and magnetic responses of the material, respectively. They can be determined as follows
\begin{eqnarray}
\sigma_{sca}^{\text{P}} = \frac{\pi}{k^2} \sum_{l = 1}^{ \infty } \sum_{m = -l}^{l} \left( 2 l + 1  \right) \left(  \vert a_{E}^{P} (l,m) \vert^2 +  \vert a_{H}^{P} (l,m) \vert^2  \right), \\
\sigma_{sca}^{\text{M}} = \frac{\pi}{k^2} \sum_{l = 1}^{ \infty } \sum_{m = -l}^{l} \left( 2 l  + 1  \right) \left(  \vert a_{E}^{M} (l,m) \vert^2 +  \vert a_{H}^{M} (l,m) \vert^2  \right).
\end{eqnarray}

The term $\Re\{ \sigma_{sca}^{\text{PM}}\}$ originates from the interaction between the scattered electromagnetic waves produced by the two types of bound currents ---i.e., the polarization and magnetization current densities--- and is given by
\begin{eqnarray}
 \sigma_{sca}^{\text{PM}}  = \frac{\pi}{k^2} \sum_{l = 1}^{ \infty } \sum_{m = -l}^{l}  \left( 2 l + 1  \right) \left( a_{E}^{P} (l,m) \overline{a_{E}^{M} (l,m)} + a_{H}^{P} (l,m) \overline{a_{H}^{M} (l,m)} \right).
\end{eqnarray}
 
The extinction power $ W_{ext}$ is determined using the angular components of the scattered electromagnetic fields,\cite{bohren1998}
\begin{eqnarray}
W_{ext}  = - \frac{1}{2} \Re  \bigg\{\int_{\omega} \left[  E_{i \theta} H^{*}_{s \varphi} -  E_{i \varphi} H^{*}_{s \theta} \right.  + \left. E_{s \theta} H^{*}_{i \varphi} -  E_{s \varphi} H^{*}_{i \theta} \right] r^2 \; d \Omega\bigg\} .
\end{eqnarray}

Although the scattering cross section is independent of the polarization of the incident plane wave, the extinction cross section depends on the polarization of the incident electric field. Hence, the extinction cross sections for a plane wave with the electric field polarized along the $x$- and $y$-axis are
\begin{eqnarray}
&& \sigma_{ext, x} = \sigma_{ext, x}^{\text{P}} + \sigma_{ext, x}^{\text{M}}, \\
&& \sigma_{ext, y} = \sigma_{ext, y}^{\text{P}} + \sigma_{ext, y}^{\text{M}} .
\end{eqnarray}
As observed, the last expressions do not include the electric-magnetic (PM) cross terms since extinction is formulated in terms of the products between the angular components of the scattered fields and the incident light. For an incident wave with the electric field polarized along the $x$-axis, the magnetic component of the extinction cross section can be computed as
\begin{eqnarray}
\sigma_{ext, x}^{\text{M}} = - \frac{\pi}{k^2} \sum_{l = 1}^{ \infty } \sum_{m = -1,1} \left( 2 l +  1  \right)  \Re \{  m a_E^{\text{M}} (l,m) + a_H^{\text{M}} (l,m) \},
\end{eqnarray}
The contribution from the dielectric polarization density can be obtained by replacing M with P. Similarly, for an incident plane wave with the electric field polarized along the 
$y$-axis, the magnetic component of the extinction cross section is 
\begin{eqnarray}
\sigma_{ext, y}^{\text{M}} = \frac{\pi}{k^2} \sum_{l = 1}^{ \infty } \sum_{m = -1,1} \left( 2 l  +  1  \right)  \Im \{  a_E^{\text{M}} (l,m) + m a_H^{\text{M}} (l,m) \}.
\end{eqnarray}
Again, the contribution from the dielectric polarization density can be obtained by replacing M with P. On the other hand, the absorbed power can be calculated as \( \sigma_{abs} = \sigma_{ext} - \sigma_{sca} \). In the following sections, numerical results are presented in terms of the scattering ($ Q_{sca} $), extinction ($ Q_{ext} $), and absorption ($ Q_{abs} $) efficiency coefficients, which represent the cross sections normalized by the particle’s projected area.

\section{Numerical Assessment of the Analytical Model}

For completeness of the numerical analysis, we compared in the Appendix the contributions to efficiency from each multipole mode for spheres of different materials with the results from Mie theory (multipole by multipole). In particular, we compared the scattering cross sections of isotropic dielectric–magnetic spheres with those predicted by Mie theory. Furthermore, we performed a near-field decomposition of the electric field to analyze the individual contributions of multipolar modes to the polarization and magnetization densities, considering terms up to multipole order $ l=4 $ (i.e., up to the hexadecapolar terms). The results using our formalism show excellent agreement with conventional Mie theory predictions, validating the accuracy of the approach. This type of analysis illustrates the advantages of spherical multipole decomposition methods, which express the scattered field in terms of vector spherical harmonics. Unlike other standard multipole approaches ---typically constrained to the far-field region and limited to evaluating far-field observables--- our theoretical framework enables a rigorous decomposition of electromagnetic fields in the reactive near-field and the radiating near-field (Fresnel) zones as well. The ability to perform a multipole decomposition that is exact, complete, and valid across all three spatial regions surrounding a scatterer is essential for numerous advanced applications. These include, for example, the multipolar analysis of electromagnetic chirality density in the near field,\cite{Mie_sphere_acsphotonics} and the study of near-field radiative heat transfer (NFRHT) between closely spaced scatterers at the multipolar level.\cite{Edwin_Cuevas_PhysRevB} The corresponding results are discussed in the Appendix section.

\begin{figure}[h!]
\centering
\includegraphics[scale=0.5]{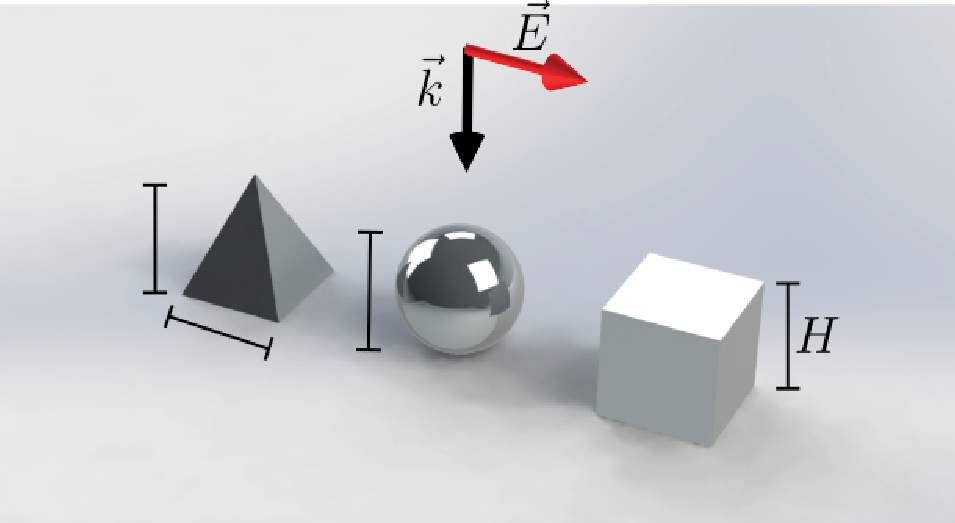}
\caption{The scatterers considered are millimeter-scale structures with constant material properties. The three geometries analyzed include a pyramid, a sphere, and a cube. For each structure, the defining geometric parameters are as follows: the edge length of the cube, the diameter of the sphere, and both the height and base length of the pyramid are all set to $H=65$~mm. These particles are illuminated by a monochromatic plane wave propagating along the $z$-axis, with the electric field $\mathbf{E}$ polarized along the $x$-axis.}
\label{fig_1}
\end{figure}

We shall later compare the numerical results from our model with full-wave simulations using COMSOL Multiphysics. This commercial software is recognized for its robust electromagnetic simulation capabilities in photonics applications from microwave to optical regimes. To further demonstrate the reliability of the approach, we performed calculations for three distinct geometries ---pyramidal, spherical, and cubic--- as illustrated in Figure~\ref{fig_1}. Since the primary focus is on far-field characteristics, we present comparative results for the scattering, extinction, and absorption cross-sections as functions of the wavelength. For clarity and consistency, the simulations assume constant material properties for the scatterers under a time-harmonic electromagnetic field. 

\begin{figure}[t]
\includegraphics[width=1\columnwidth]{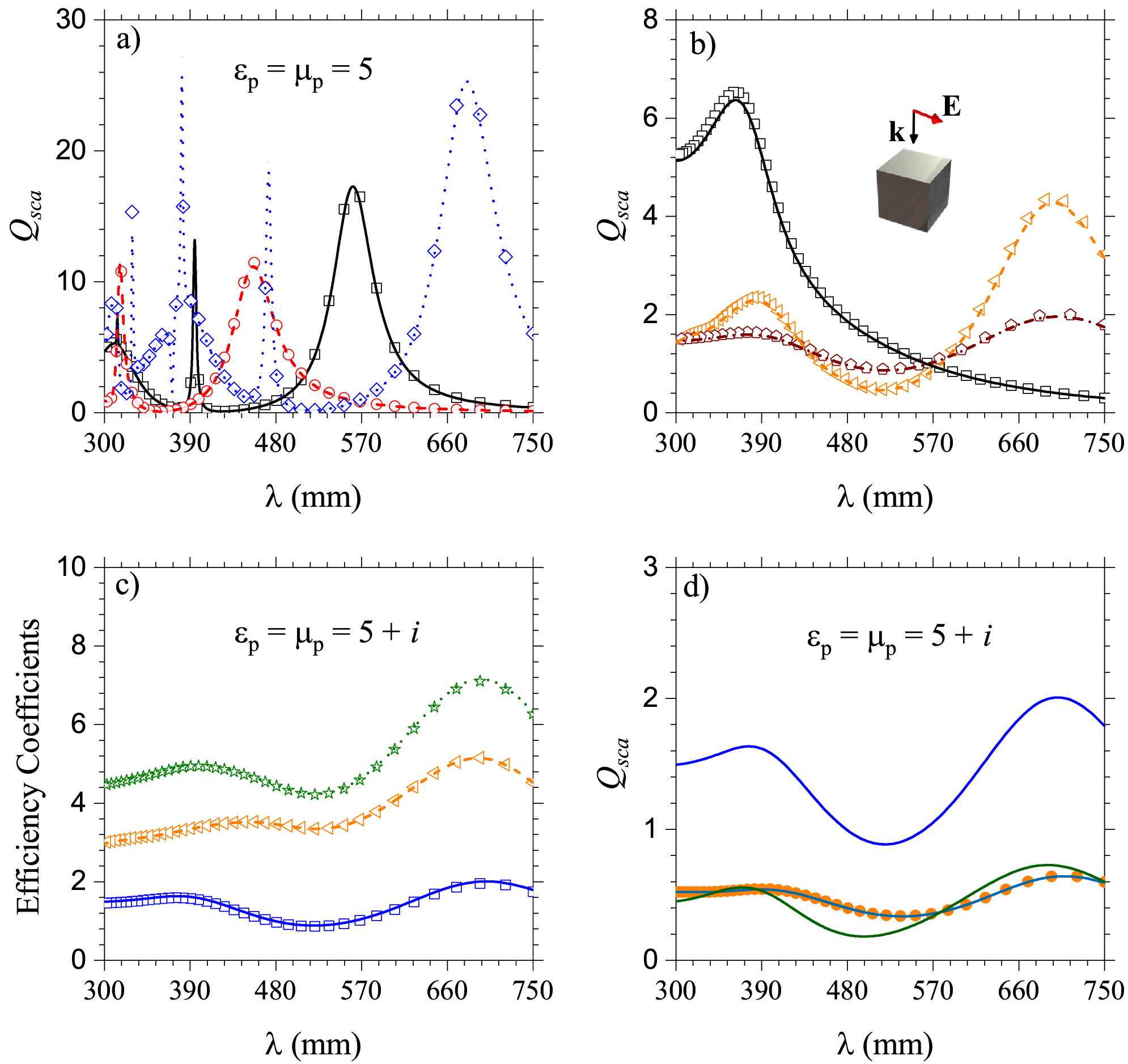}
\caption{The scattering efficiency coefficient and corresponding multipole decomposition contributions were calculated for (a) three distinct particles: a sphere, a cube, and a pyramid, and (b) for various scalar relative permittivity and permeability values for the cube. In detail, for a cube with material parameters \( \varepsilon_p = \mu_p = 5 + i \), the extinction (\( Q_{\text{ext}} \)), scattering (\( Q_{\text{scat}} \)), and absorption (\( Q_{\text{abs}} \)) efficiency coefficients are presented in (c), while the scattering cross sections attributed to polarization, magnetization, interaction between the EM fields generated by both bound currents and the total contribution are illustrated in (d). The particles are considered to be immersed in a vacuum (\( \varepsilon_s = \mu_1 = 1 \)) and irradiated by a linearly polarized plane wave incident from above.}
\label{fig2}
\end{figure}

Figure~\ref{fig2} compares the numerical results obtained from the analytical expressions with full-wave electromagnetic simulations in COMSOL Multiphysics. Specifically, Figure~\ref{fig2} (a) illustrates the scattering efficiency $ Q_{sca} $ for the three geometries considered. The analytical results are represented by black solid (sphere), red dashed (pyramid), and blue dotted lines (cube), while the corresponding COMSOL simulations are depicted using open squares, open circles, and open diamonds, respectively. In particular, COMSOL Multiphysics enables the calculation of total scattering efficiency by integrating the time-averaged Poynting vector over a spherical surface enclosing the scatterer. However, this approach does not permit the identification or separation of the individual multipole contributions excited within the magneto-dielectric structure. In contrast, the theoretical framework developed in this work not only allows for the computation of the total scattering efficiency ---by summing the contributions of all multipole modes--- but also enables the explicit evaluation of the scattering efficiency associated with each multipole order. All calculations were performed for dielectric particles with permittivity and permeability values of $\varepsilon=\mu=5$. To further validate the approach and investigate the effects of variations in dielectric material properties, we analyzed different combinations of $(\varepsilon,\mu)$ values for the cubic geometry, as shown in Figure~\ref{fig2}(b). In this latter figure, the solid black line represents $(1,5)$, the orange dashed line denotes $(5,5+i)$, and the wine dash-dotted line indicates $(5+i,5+i)$. The corresponding COMSOL simulations are represented by open diamonds, open squares, open triangles, and open pentagons, respectively. The excellent agreement between the analytical approach and full-wave numerical simulations in COMSOL Multiphysics demonstrates the reliability of the methodology used here. This validation confirms that the approach can model the scattering properties of dielectric-magnetic particles with arbitrary geometries and material properties.

Having validated the approach for $ Q_{sca} $ across different geometrical and material parameters, we now verify the extinction efficiency $ Q_{ext} $ and absorption efficiency $ Q_{abs} $  as well. Figure~\ref{fig2}(c) shows the efficiency coefficients, where $ Q_{ext} $, $ Q_{sca} $, and $ Q_{abs} $ are denoted by green dotted, orange dashed, and blue solid lines, respectively. The corresponding COMSOL simulations are depicted using open stars, open triangles, and open squares, demonstrating excellent agreement with the analytical results. 

Finally, to highlight one of the key advantages of the analytical approach over COMSOL simulations, Figure~\ref{fig2}(d) illustrates the decomposition of $ Q_{sca} $ (calculated for the cube with $\varepsilon_p = \mu_p = 5 + i $), represented by a blue solid line, into its individual components: $ Q_{sca}^{\text{P}}$ (blue solid line), $ Q_{sca}^{\text{M}}$ (orange solid circles), and $ Q_{sca}^{\text{PM}}$ (green solid line). This detailed decomposition provides deeper physical insight into the far-field properties of the electromagnetic waves generated due to polarization density, magnetization density and the interaction between both waves.

\section{MAGNETIC RESONANCE IN FERRITE CYLINDERS}
\begin{figure}[t]
\includegraphics[width=1\linewidth]{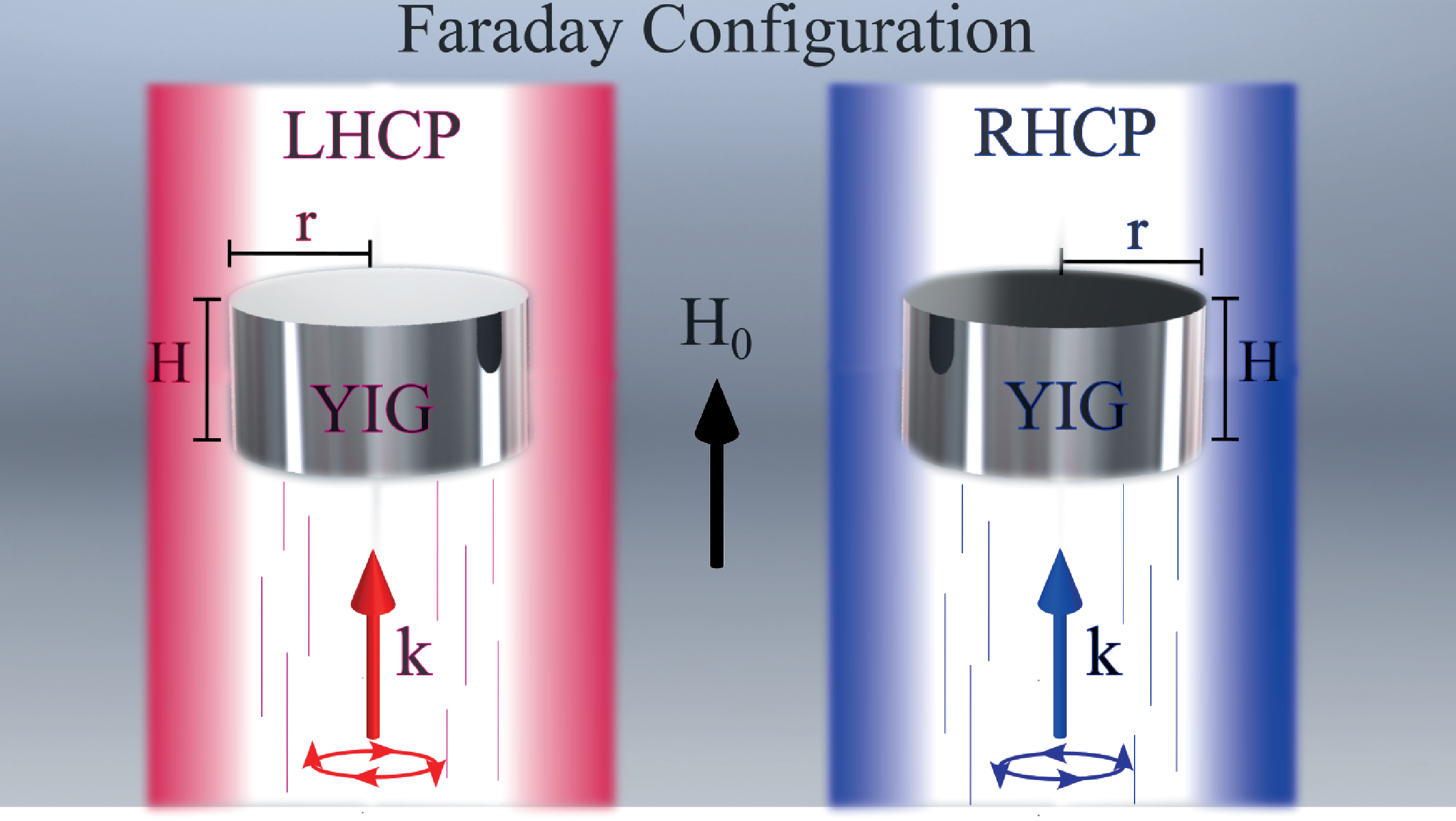}
\caption{This is a graphical representation of a ferrite cylinder (yttrium iron garnet (YIG)) with height \( H \) and radius \( r \), positioned in an external static magnetic field \( H_0 \), oriented in the Faraday configuration, meaning it is parallel to the propagation direction of the incident radiation (\( \mathbf{H}_0 \parallel \mathbf{k} \)). The cylinder is excited by incident waves from below, specifically by (a) left-hand circularly polarized (LHCP) and (b) right-hand circularly polarized (RHCP) waves.} 
\label{fig3}
\end{figure}

The practical relevance of our analytical findings is demonstrated by examining the far-field properties of a ferrite millimeter-scale cylinder under an external static magnetic field, in addition to their interaction with electromagnetic millimeter waves (mmWaves). In this setup, a static magnetic field $\mathbf{H}_0 $ ---which may represent either the intrinsic magnetization of the ferrite material or an externally applied field--- is aligned with the wave vector $\textbf{k}$ of the incident mmWave, establishing the Faraday configuration. According to the Faraday effect, electromagnetic waves propagating parallel to a static magnetic field undergo either right-hand circular polarization (RHCP) or left-hand circular polarization (LHCP) rotations, depending on whether $\textbf{k}$ is aligned parallel or antiparallel to $\mathbf{H}_0 $.\cite{pozar_2012} To broaden the applicability of the approach, we now consider circularly polarized (CP) electromagnetic waves to analyze the far-field properties of LHCP and RHCP waves after interacting with a magnetized cylinder, as illustrated in Figure~\ref{fig3}. This approach allows us to assess the feasibility of determining the polarization-dependent wave behavior in magnetized ferrite structures. For experimental feasibility, we model the millimeter-scale cylinders using yttrium iron garnet (YIG) and conduct simulations based on experimentally validated parameters. The external static magnetic field strength is set to \( H_0 = 278520 \, \mathrm{A/m} \); with a susceptibility linewidth near resonance of \( \Delta H = 3581 \, \mathrm{A/m} \) and a saturation magnetization of \( M_0 = 141650 \, \mathrm{A/m} \). The material's relative permittivity is \( \varepsilon_p = 15 \), and the loss tangent is \( \tan \delta = 0.0002 \).\cite{pozar_2012} These parameters provide a realistic basis for investigating the far-field electromagnetic response of magnetized ferrite cylinders in the mmWave regime. The mathematical expressions for the elements of the anisotropic permeability tensor of ferrite are provided in the Appendix section.

\begin{center}\
\begin{figure}[h]
\includegraphics[width=\linewidth]{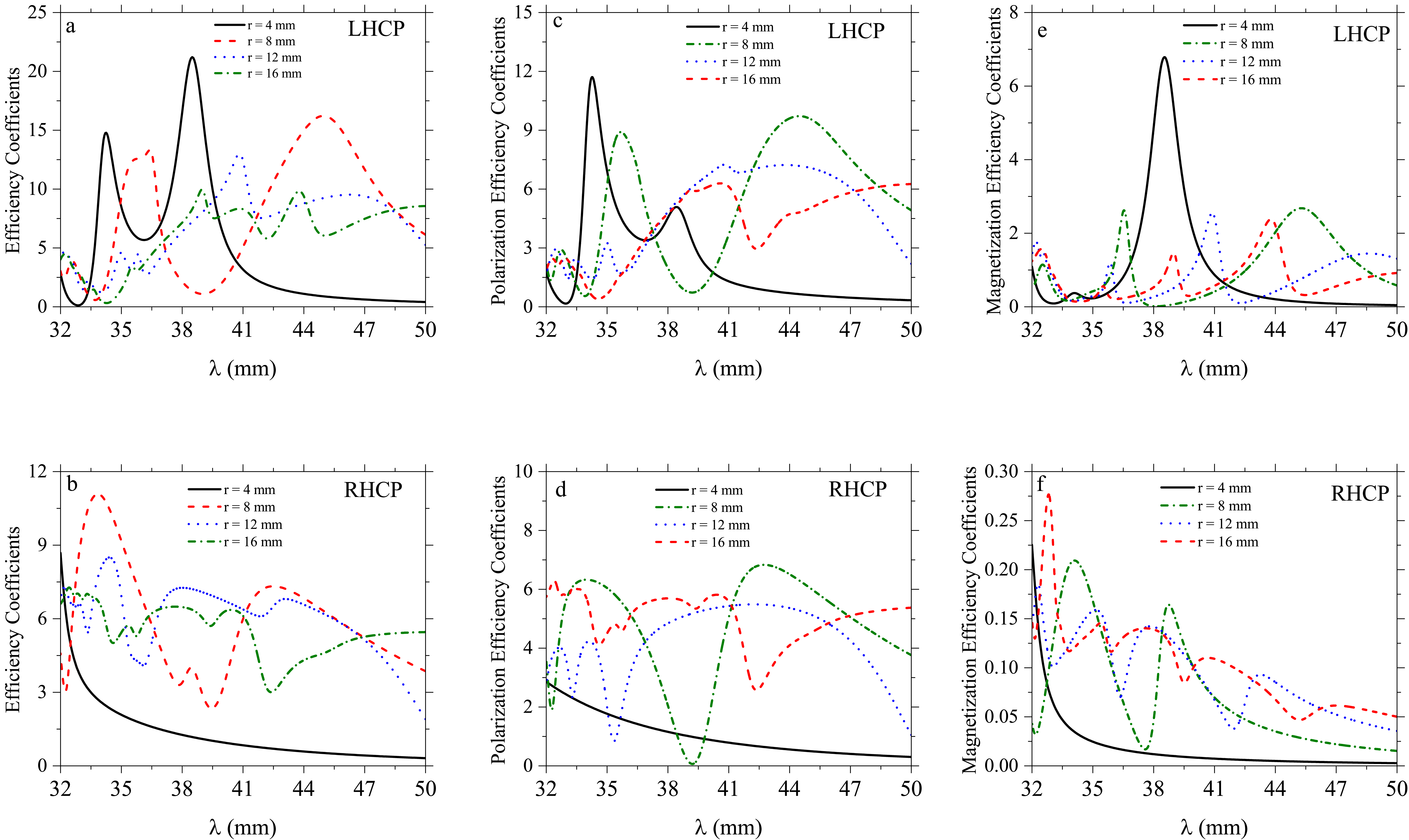}
\caption{\label{fig4}The total scattering efficiency coefficients (\( Q_{\text{scat}} \)) and their respective contributions from polarization and magnetization currents are shown for a ferrite cylinder with varying radii ($ r = 4, 8, 12, 16 $ mm). The cylinder is irradiated by circularly polarized waves in the Faraday configuration: (a), (c), and (e) depict results for left-hand circularly polarized (LHCP) waves, while (b), (d), and (f) correspond to right-hand circularly polarized (RHCP) waves.}
\end{figure}\
\end{center}

\begin{figure}[t]
\includegraphics[width=\columnwidth]{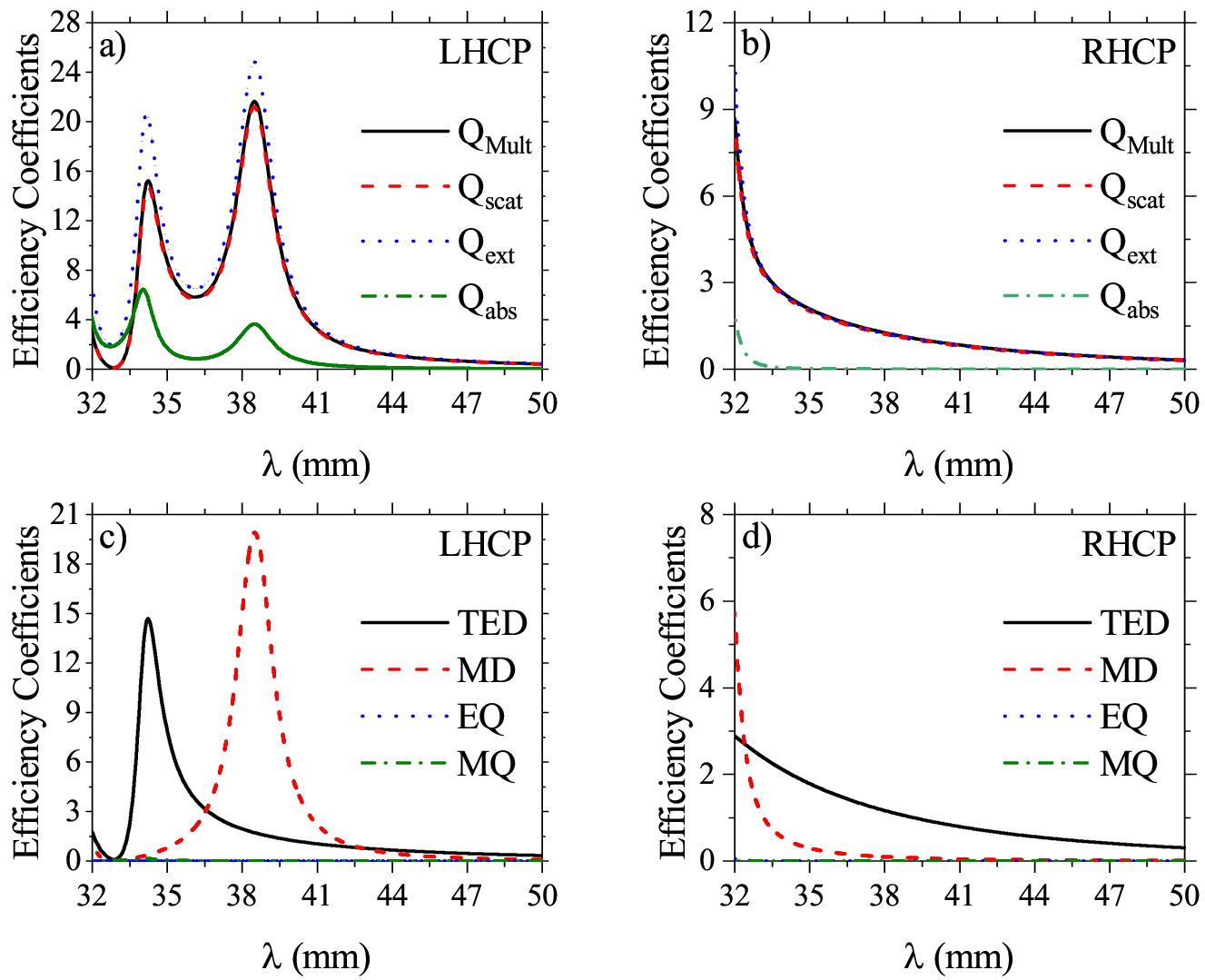}
\caption{\label{fig5}The extinction (\( Q_{\text{ext}} \)), scattering (\( Q_{\text{scat}} \)), and absorption (\( Q_{\text{abs}} \)) efficiency coefficients, along with the multipole decomposition of the scattering efficiency coefficient (\( Q_{\text{Mult}} \)), are analyzed for a ferrite cylinder irradiated by (a) left-hand circularly polarized (LHCP) and (b) right-hand circularly polarized (RHCP) waves in the Faraday configuration. The multipole decomposition contributions up to the fourth order, including dipole and quadrupole terms, are shown for (c) LHCP and (d) RHCP cases.}
\end{figure}

Figures~\ref{fig4}(a)-\ref{fig4}(c) and \ref{fig4}(d)-\ref{fig4}(f) present numerical results for the scattering cross-section corresponding to LHCP and RHCP waves, respectively. The results are shown for cylinders with varying radii while maintaining a constant height of $H= 4 $~mm. The scattering cross-section was computed using the sum of the first four multipoles. Figures~\ref{fig4}(a) and \ref{fig4}(d) display the total scattering cross-section in excellent agreement with COMSOL Multiphysics simulations, which is further illustrated in Figures~\ref{fig5}(a) and \ref{fig5}(b). \( Q_{\text{scat}} \), \( Q_{\text{ext}} \) and \( Q_{\text{abs}} \) were obtained using COMSOL Multiphysics, while the term \( Q_{\text{Mult}} \) represents the scattering efficiency calculated using the mathematical expression derived in this work. To distinguish the individual contributions of currents associated with \textbf{P} and \textbf{M}, we present comparative calculations of the scattering  efficiencies. Figures~\ref{fig4}(b) and \ref{fig4}(e) show results excluding \textbf{M}, while Figures~\ref{fig4}(c) and \ref{fig4}(f) display results without \textbf{P}, for LHCP and RHCP polarizations. These results highlight the necessity of incorporating magnetic contributions into the spherical multipole expansion to achieve an accurate description of the far-field properties of ferrites under an external static magnetic field. Specifically, the resonant peaks observed at $r=4$ in Figure~\ref{fig4}(a) (and correspondingly in Figure~\ref{fig4}(d)) can only be explained properly by combining the contributions from Figures~\ref{fig4}(b) and \ref{fig4}(c) (or Figures~\ref{fig4}(e) and \ref{fig4}(f), respectively). It is worth noting that this superposition is mentioned solely for visualization purposes. An exact reproduction of the total scattering results require accounting for the cross-term contribution between \textbf{P} and \textbf{M}, denoted with super-index PM in Eq. (\ref{totalscaPM}). 

Next, we decompose \( Q_{\text{Mult}} \) from Figures~\ref{fig5}(a)-\ref{fig5}(b) (associated with $r=4$~mm) into its constituent multipole components. The results in Figures~\ref{fig5}(c)-\ref{fig5}(d) illustrate the contributions of the total electric dipole (TED), magnetic dipole (MD), electric quadrupole (EQ), and magnetic quadrupole (MQ) terms to \( Q_{\text{Mult}} \) for LHCP and RHCP polarized waves, respectively. Figure~\ref{fig5}(c) reinforces the observations from Figure~\ref{fig4}, emphasizing the crucial role of magnetic density currents in multipole calculations. Their inclusion is essential for representing the far-field characteristics of magnetic scatterers accurately. Notably, Figure~\ref{fig5}(c) reveals that the far-field behavior of the ferrite resonator is primarily governed by the interaction of two strong radiating dipoles ---one electric and one magnetic. The differences in far-field behavior between LHCP and RHCP waves, as observed in Figures~\ref{fig5}(c)-\ref{fig5}(d), stem from the opposite Faraday rotations induced by the external magnetic field $\mathbf{H}_0 $, which is oriented along the $z$-axis as indicated in Figure~\ref{fig3}. 

\begin{figure}[t!]
\includegraphics[width=\columnwidth]{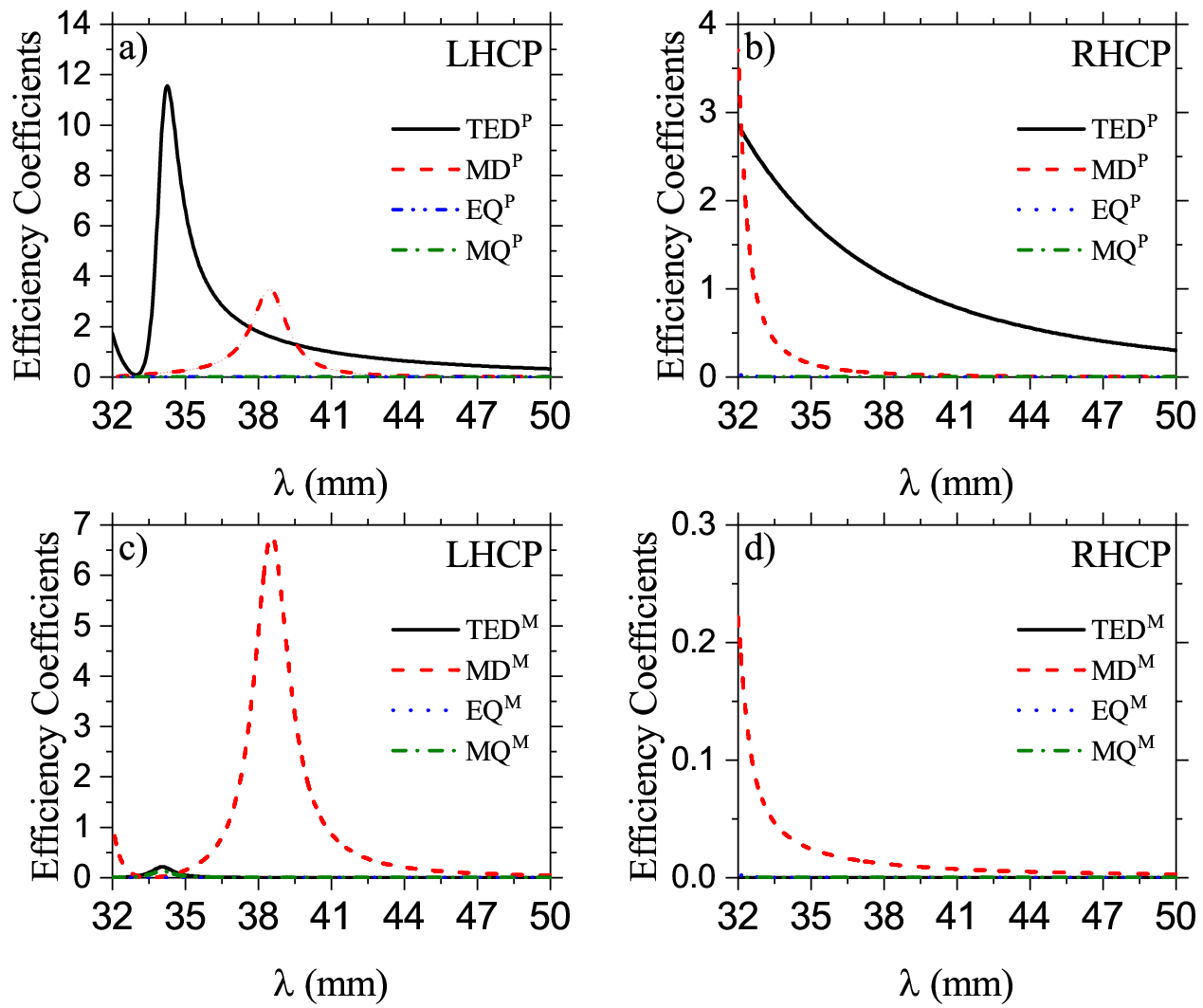}
\caption{\label{fig6}Separate contributions from the polarization and magnetization current densities to the scattering efficiency coefficients of a ferrite cylinder. The multipole decomposition of the polarization density, representing the redistribution of bound charges, is shown in (a) for LHCP light and in (b) for RHCP light. Similarly, the contribution from the magnetization density arising from the unpaired spins is shown in (c) for LHCP light and in (d) for RHCP light.}
\end{figure}

It has been demonstrated that magnetic resonances can be excited using precisely engineered all-dielectric, nonmagnetic resonant structures~\cite{Evlyukhin2012,Zhang2019,Tuz2020}. The analytical framework developed here now allows one to determine how the far-field properties of all-dielectric magnetic resonators are governed by magnetic current densities. These magnetic currents are often neglected but they are essential to achieve a precise, and accurate characterization of all-dietectric magnetic systems. To illustrate this point, Figures~\ref{fig6}(a)-(b) show the numerical results from Figures~\ref{fig5}(c)-\ref{fig5}(d) while omitting the contribution of \textbf{M}. Figures~\ref{fig6}(c)-(d) show the same results while neglecting the contribution of \textbf{P}. These comparisons reveal a crucial insight: although a weak MD can be excited solely by \textbf{P} (as expected for all-dielectric, nonmagnetic resonators), the dominant magnetic dipole resonance observed in the scattering efficiency of all-dielectric magnetic resonators can only be reproduced accurately when the contribution of \textbf{M} is considered.

\section{Multipole Decomposition of Magnetic Circular Dichroism}

\begin{figure}[t!]
\includegraphics[width=\columnwidth]{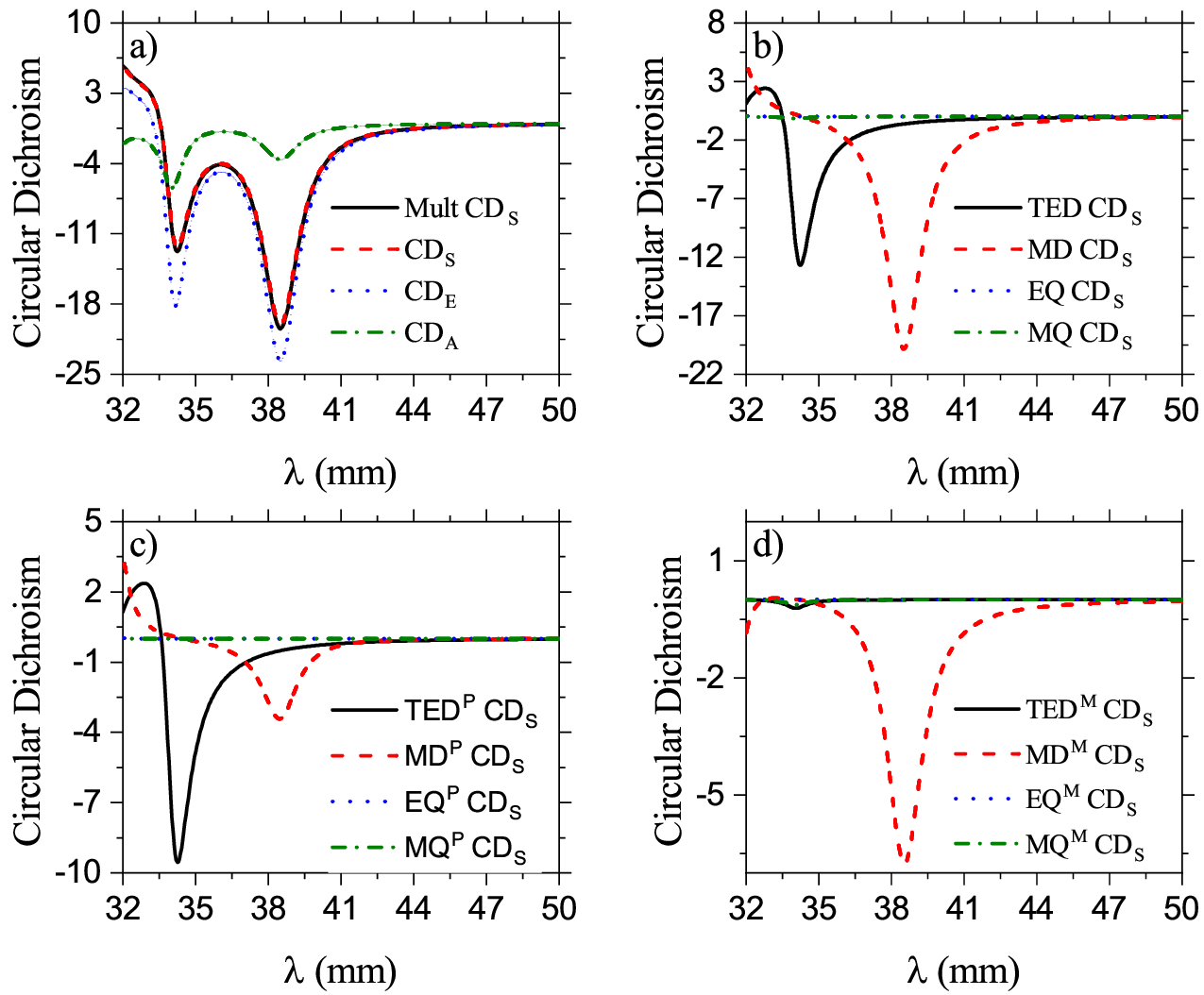}
\caption{\label{fig7}Figure (a) presents the circular dichroism in dispersion, extinction and absorption for a ferrite cylinder in the Faraday configuration. In Figure (b), the contributions from the first four multipoles (dipoles and quadrupoles) to the circular dichroism are shown. Figure (c) shows the contributions of the circular dichroism from the polarization density due to the first four multipoles, while Figure (d) illustrates the corresponding contributions from the magnetization density.}
\end{figure}

The presence of $\mathbf{H}_0 $, as illustrated in Figure~\ref{fig3}, imparts handedness to the studied cylinder. This magnetically induced chirality, in turn, leads to preferential absorption and scattering of CP electromagnetic waves, either LHCP or RHCP. Since the most established spectroscopic technique for measuring this phenomenon is magnetic circular dichroism (MCD), which quantifies the differential absorption and scattering of LHCP and RHCP waves, we calculated CD for the extinction (\( \text{CD}_{\text{E}} = Q_{\text{Ext}}^{\text{RHCP}} - Q_{\text{Ext}}^{\text{LHCP}} \)), scattering (\( \text{CD}_{\text{S}} = Q_{Sca}^{\text{RHCP}} - Q_{Sca}^{\text{LHCP}} \)), and absorption (\( \text{CD}_{\text{A}} = Q_{Abs}^{\text{RHCP}} - Q_{Abs}^{\text{LHCP}} \)),\cite{Zhu_2023} as shown in Figure~\ref{fig7}(a). For validation, the results in Figure~\ref{fig7}(a) were obtained using COMSOL Multiphysics for $\text{CD}_{\text{E}}$, $\text{CD}_{\text{S}}$ and $\text{CD}_{\text{A}}$. Additionally, for visualization purposes, the results from the multipole method are presented comparatively only for $\text{CD}_{\text{S}}$ (labeled Mult\_$\text{CD}_{\text{S}}$), demonstrating excellent agreement between both approaches. Figure~\ref{fig7}(b) shows the contribution from first four individual multipole components to $\text{CD}_{\text{S}}$. Only the two lowest-order terms exhibit significant contributions in this case. The contributions from \textbf{P} and \textbf{M} are shown in Figures~\ref{fig7}(c) and~\ref{fig7}(d), respectively, again emphasizing the importance of including \textbf{M} to provide a precise, complete explanation of electromagnetic far-field measurements from magnetic scatterers.

\section{Conclusions}

We have investigated numerically the scattering properties of dielectric-magnetic particles with arbitrary shapes, characterized by tensorial permeability and permittivity. In particular, we demonstrated that the multipole decomposition of magnetic circular dichroism from magnetized objects offers versatility for gaining deeper insights into the far-field behavior of dielectric-magnetic scatterers, especially for nanophotonics and radio frequency systems. Use was made of the exact expressions for the electric and magnetic spherical multipole coefficients of wave scattering by magnetic particles. Furthermore, we present a full multipole decomposition of the scattered fields and scattering efficiencies that is valid up to any order using vector spherical harmonics, providing a more comprehensive understanding of scattering phenomena. Numerical results were validated through comparison with Mie theory and COMSOL Multiphysics simulations using the finite element method, confirming the accuracy and reliability of the analytical framework in modeling dielectric-magnetic particles with complex shapes and material properties. These findings emphasize the critical role of magnetization currents in achieving a precise understanding of far-field features in dielectric-magnetic resonators. Finally, this study can be extended to analyze 2D dielectric-magnetic metasurfaces with arbitrary unit cell shapes up to any order.\cite{Dezert_2019, Jessica_2024_TAP} Such analysis will help identify magnetization current contributions to transmission and reflection coefficients, thus making it possible to design advanced photonic devices.

\section{Appendix}

\section*{Mathematical Simplifications Using Integration by Parts}

In this section, we present the intermediate calculations used in the integration by parts, which was performed to circumvent the difficulties associated with computing derivatives of numerical quantities, specifically the polarization and magnetization densities.\cite{Grahn2012} Below, we demonstrate how each of the integrals involving the magnetization density is rewritten. 

The first integral represents the contribution of the magnetization density to the electric spherical multipole coefficients and is given by  
\begin{eqnarray*}
\int Y_{lm}^{*} j_{l} \left( kr \right) \left[ \mathbf{\nabla} \cdot \left( \mathbf{r} \times \mathbf{M} \right) \right] d^3 r  =  \Xi_{lm} \int_{V} e^{-i m \varphi} j_l \left( kr \right) \left\lbrace i \pi_{lm} \left( \theta \right) \hat{\theta} \cdot \mathbf{M} + \tau_{lm} \left( \theta \right) \hat{\varphi} \cdot \mathbf{M} \right\rbrace  \; d^{3} r .
\end{eqnarray*}

Notably, only the angular components of the magnetization density contribute to the electric spherical multipole coefficients. The next three terms correspond to contributions of the magnetization density to the magnetic spherical multipole coefficients:
\begin{equation}
\frac{1}{k^2} \int Y_{lm}^{*} j_{l}\left( kr \right) \nabla^2 \left(  \mathbf{r} \cdot \mathbf{M} \right) d^3 r  = - \frac{\Xi_{lm}}{k}  \int_{V} e^{-i m \varphi} \Psi_l \left( kr \right) P^{m}_{l} \left( \cos \theta \right) \hat{r} \cdot \mathbf{M} \; d^{3} r \; .
\end{equation}

It is important to note that, in the main article, Green’s theorem was employed to replace $ \nabla^2 $ by $ - k^2 $. The contribution of this term arises from the radial component of the magnetization density. Consequently, the second integral involved in computing the electric spherical multipole coefficients can be rewritten using integration by parts as follows:
\begin{eqnarray}
\nonumber
& & - \int Y_{lm}^{*} j_{l}\left( kr \right) \left( \mathbf{\nabla} \cdot \mathbf{M} \right) d^3 r  = \int_{V} Y^{*}_{lm} \frac{\partial}{\partial r} j_l \left( kr \right) \hat{r} \cdot \mathbf{M} \; d^{3} r  \\ \nonumber
& & + \Xi_{lm} \int_{V} e^{-i m \varphi} \frac{j_l \left( kr \right)}{r} \left\lbrace  \tau_{lm} \left( \theta \right)  \hat{\theta} \cdot \mathbf{M} - i \pi_{lm} \left( \theta \right) \hat{\varphi} \cdot \mathbf{M} \right\rbrace \; d^{3} r \; .
\end{eqnarray}

Finally, the last term is given by  
\begin{eqnarray}
\nonumber
& & \int Y_{lm}^{*} j_{l} \left( kr \right) r \frac{\partial}{\partial r} \left( \mathbf{\nabla} \cdot \mathbf{M} \right) d^3 r  = \int_{V} Y^{*}_{lm} \left[ 2 \frac{\partial}{\partial r} j_l \left( kr \right) + \frac{\partial^2}{\partial r^2} \left( r j_{l} \left( kr \right) \right) \right] \hat{r} \cdot \mathbf{M}  \; d^3 r . \\
& & +  \Xi_{lm} \int_{V} e^{-i m \varphi} \left[ 3 \frac{ j_l \left( kr \right)}{r}  + \frac{\partial}{\partial r} j_{l} \left( kr \right) \right] \left[ \tau_{lm} \left( \theta \right)  \hat{\theta} \cdot \mathbf{M} - i  \pi_{lm} \left( \theta \right) \hat{\varphi} \cdot \mathbf{M} \right] \; d^3 r \; .
\end{eqnarray}

To derive the corresponding contributions from the polarization density, a similar procedure must be followed. Note that all boundary terms vanish in the limit of a spherical surface that fully encloses the scatterer, since the magnetization and polarization densities are zero outside the scattering system. Finally, by grouping the above expressions as indicated by the full integrals presented in the main article, the coefficients $ a_E^M \left( l , m \right) $ and $ a_H^M \left( l , m \right) $ derived in this study are obtained.

\section*{Comparison with Mie Theory: Scattering Cross Sections of Isotropic Magneto-Dielectric Spheres}  

The Mie solution, a rigorous analytical solution to Maxwell’s equations, describes the scattering of a plane electromagnetic wave by a homogeneous spherical particle. This solution is expressed as an infinite series of spherical multipole partial waves. Derived from Lorenz-Mie functions, it utilizes expansion coefficients of spherical functions to quantify scattering, extinction, and absorption cross sections \cite{bohren1998}. In particular, the extinction and scattering cross sections are given by:  

\begin{equation}
    \sigma_{ext} = \frac{2\pi}{k^2} \sum_{n=1}^{\infty} (2n+1) \Re \left( a_n + b_n \right),
\end{equation}

\begin{equation}
    \sigma_{sca} = \frac{2\pi}{k^2} \sum_{n=1}^{\infty} (2n+1) \left( \left| a_n \right|^2 + \left| b_n \right|^2 \right),
\end{equation}  

where \( a_n \) and \( b_n \) are the electric and magnetic multipole coefficients, respectively, and \( k \) is the wavenumber of the surrounding medium. The absorption cross section (\( \sigma_{abs} \)) is obtained as the difference between the extinction and scattering cross sections:  

\begin{equation}
    \sigma_{abs} = \sigma_{ext} - \sigma_{sca}.
\end{equation}  

These cross sections depend on the particle size parameter, defined as \( x = 2\pi a/\lambda \), where \( a \) is the particle radius and \( \lambda \) is the wavelength of the incident light. The multipole expansion coefficients are given by:  

\begin{equation}
    a_n = \frac{\mu m^2 j_n(mx) \left[ x j_n(x) \right]' - \mu_1 j_n(x) \left[ mx j_n(mx) \right]'}
    {\mu m^2 j_n(mx) \left[ x h_n^{(1)}(x) \right]' - \mu_1 h_n^{(1)}(x) \left[ mx j_n(mx) \right]'},
\end{equation}

\begin{equation}
    b_n = \frac{\mu_1 j_n(mx) \left[ x j_n(x) \right]' - \mu j_n(x) \left[ mx j_n(mx) \right]'}
    {\mu_1 j_n(mx) \left[ x h_n^{(1)}(x) \right]' - \mu h_n^{(1)}(x) \left[ mx j_n(mx) \right]'}.
\end{equation}  

Here, \( \mu \) and \( \mu_1 \) denote the relative permeabilities of the surrounding medium and the particle, respectively. The relative refractive index is given by \( m = n_1 / n \), where \( n \) and \( n_1 \) are the refractive indices of the medium and the particle, respectively. Finally, \( j_n(x) \) and \( h_n^{(1)}(x) \) represent the spherical Bessel and Hankel functions of the first kind, respectively.  
\begin{figure}[H]
    \centering
    \includegraphics[width=0.85\textwidth]{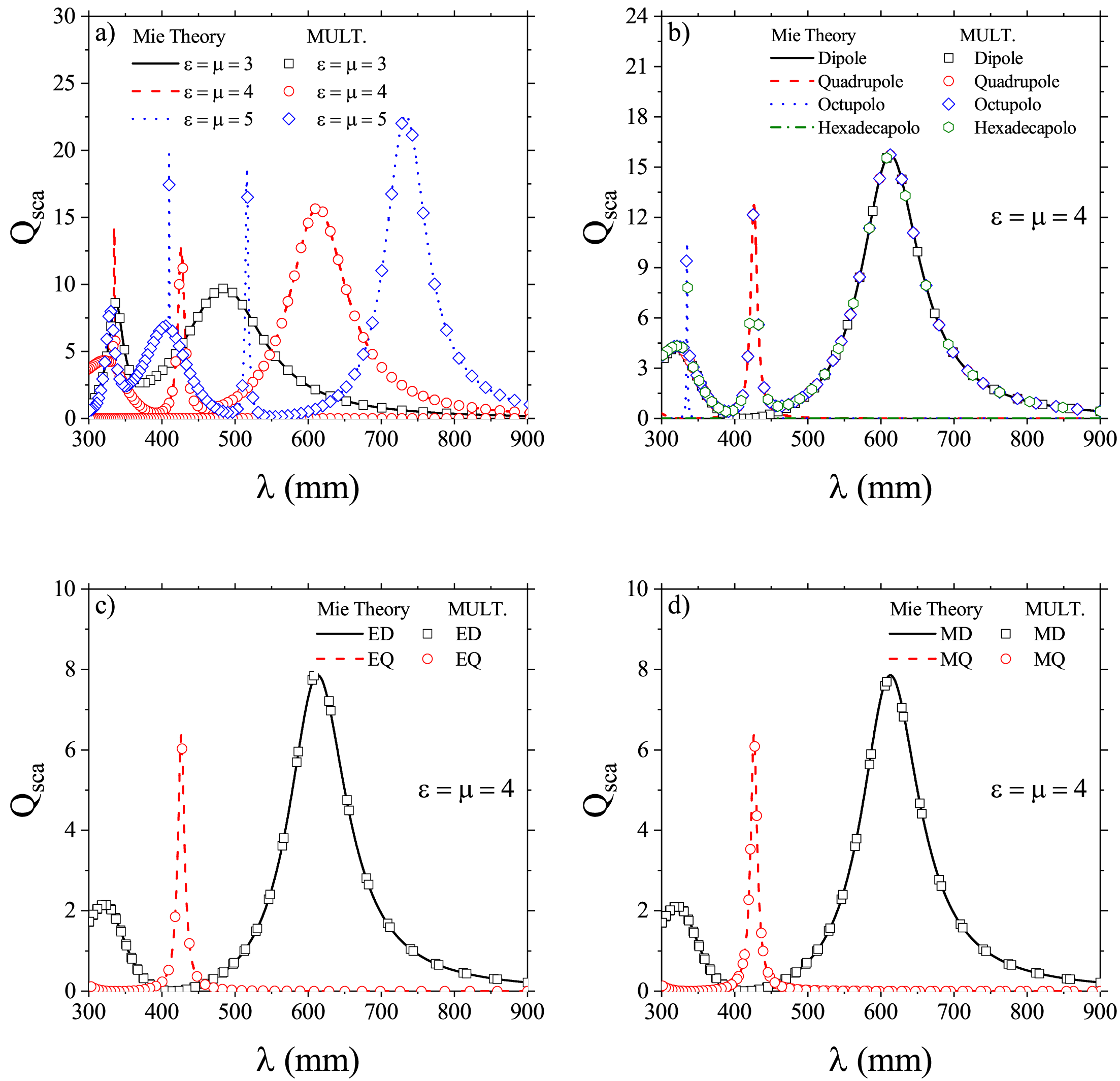}
    \caption{The scattering efficiency coefficients (\( Q_{sca} \)) and their corresponding multipole decomposition contributions were calculated using Mie theory and Spherical Multipole Coefficients (MULT) for spherical particles with a radius of 85 mm. The spheres are immersed in vacuum, i.e., \( \varepsilon_s = \mu_s = 1 \). (a) Scattering efficiencies are presented for different scalar relative permittivity and permeability values: \( \varepsilon_1 = \mu_1 = 3 \), \( \varepsilon_1 = \mu_1 = 4 \), and \( \varepsilon_1 = \mu_1 = 5 \). (b) The corresponding multipole decomposition is shown for the specific case of \( \varepsilon_1 = \mu_1 = 4 \). (c) The multipole contributions from the electric dipole and electric quadrupole moments are displayed, and (d) illustrates the multipole contributions from the magnetic dipole and magnetic quadrupole moments.}
    \label{fig:FIG1}
\end{figure}

In Figure \ref{fig:FIG1} we compare the scattering  efficients of spheres -- for a radius of $r = 85$ mm with different constant material properties, with the condition $ \varepsilon = \mu $ -- using the mathematical expressions derived in this research work with Mie's theory. This agreement extends to the spectral positions of the resonant peaks and the multipole contributions, which match their counterparts derived from the Spherical Multipole Coefficients (MULT). A numerical correspondence is observed between the contributions of electric dipole-magnetic dipole (ED-MD) and electric quadrupole-magnetic quadrupole (EQ-MQ) multipole pairs of the same order to the efficiency coefficient.\cite{kerker1983electromagnetic}

\section*{Spherical Multipole Decomposition of the Electric Near-Field Distribution of Isotropic Magneto-Dielectric Spheres}

\begin{figure}[H]
    \centering
    \includegraphics[width=12cm]{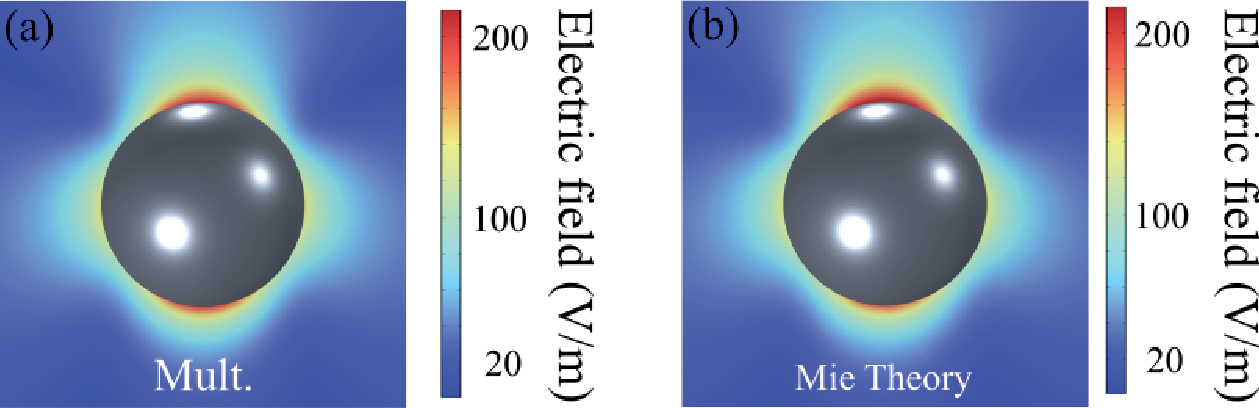}
    \caption{Scattered electric field distribution around the sphere has been calculated in (a) using the theoretical framework developed in this work. (b) same field distribution, computed using standard Mie theory. In both formalisms, multipolar contributions up to order $ l=4 $ (i.e., up to the hexadecapolar terms) were considered to ensure the accuracy of our results. The sphere has been illuminated by a monochromatic plane wave, linearly polarized along the $x$-axis and propagating along the $z$-axis, with a wavelength of 427 mm, corresponding to the quadrupole resonance. The power intensity of the incident wave is $I = 1 \, \mathrm{W/m^2}$.}
    \label{fig:FIG2}
\end{figure}

In this section, we perform a spherical multipole decomposition of the near-field electric distribution for an isotropic magneto-dielectric sphere, as this allows us to compare our results with Mie theory for each multipole field. We can thus validate the contributions of the magnetization density to the strengths of the various multiple fields, mathematical expressions derived in this work. Furthermore, recall that the spherical multipole coefficients were derived under the assumption that the scattered electromagnetic fields must be evaluated outside a spherical surface that fully encloses the scattering system. In the case of a sphere, this surface can be chosen arbitrarily close to the particle's boundary. Owing to the high symmetry of this shape, this allows for the complete reconstruction of the near-field distribution — both total and multipolar — surrounding the sphere, as is possible with Mie theory. Specifically, we consider a sphere of radius $r = 85 \, \text{mm}$, with material parameters $\varepsilon = \mu = 4$, illuminated by a linearly polarized plane wave propagating along the $z$-axis and polarized along the $x$-axis. The incident wavelength in vacuum is $\lambda = 427 \, \text{mm}$, which corresponds to a quadrupole resonance, as established in the previous section. The power intensity of the incident wave is set to $I = 1 \, \text{W/m}^2$. Figure \ref{fig:FIG2}(a) shows the scattered electric field distribution around the sphere, calculated with the theoretical framework developed in this work (note that the incident field is not included). Figure \ref{fig:FIG2}(b) presents the same field distribution, computed using standard Mie theory for validation. In both formalisms, we include multipolar contributions up to order $ l=4 $ (i.e., up to the hexadecapolar terms) to ensure the accuracy of the results.

The results in Figure  \ref{fig:FIG2} are in excellent agreement, confirming the validity of the expressions derived in our work, which account for the contributions of the magnetization density to the electric and magnetic spherical multipole coefficients. To further validate the expressions derived for the spherical multipole coefficients, and to demonstrate the scope of the multipole decomposition using the vector spherical harmonic basis, we present below the scattered electric field distributions corresponding to each of the first multipolar contributions. All multipolar field profiles are plotted in the $yz$-plane, as this plane reveals the symmetries associated with each magnetic multipole. Similar symmetry patterns can be observed for the electric multipole fields when visualized in the $xz$-plane.
\begin{figure}[H]
    \centering
    \includegraphics[width=12cm]{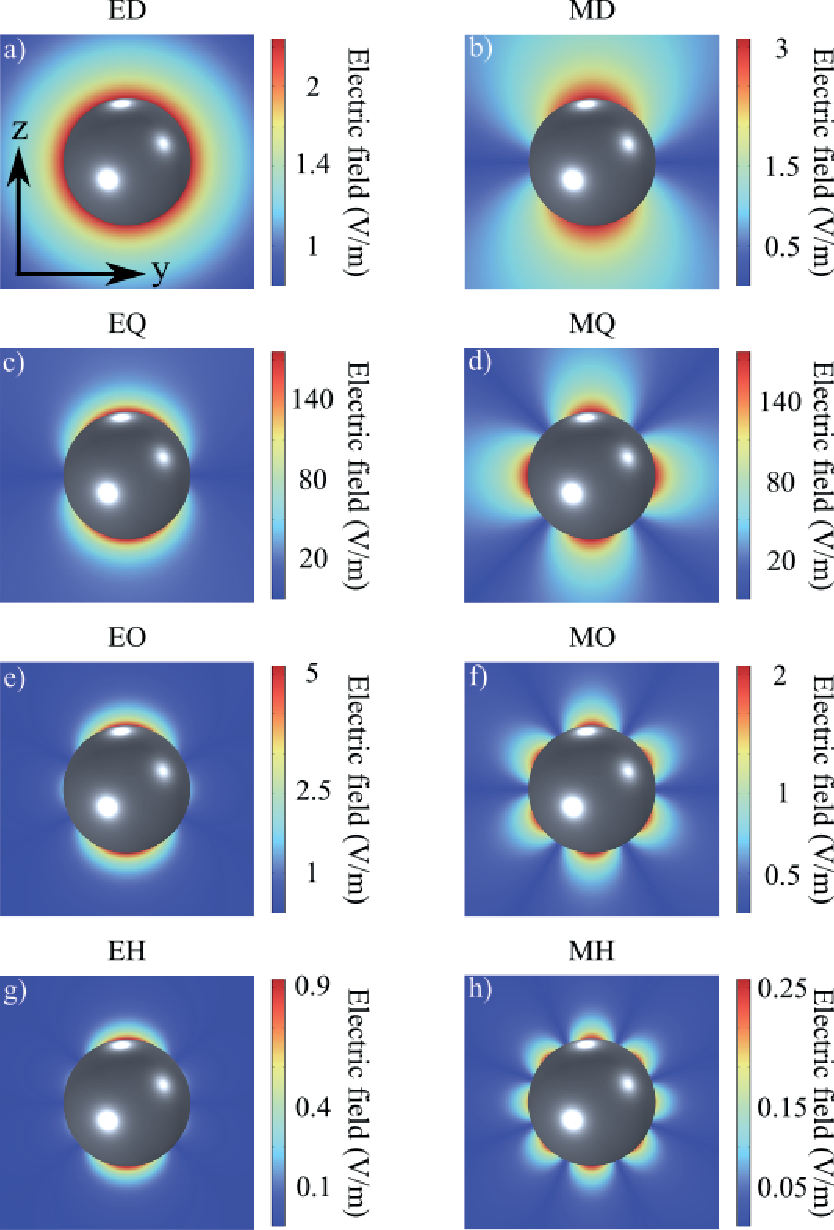}
    \caption{Multipole decomposition of the near electric field $\mathbf{E}$ in the $yz$-plane of an isotropic dielectric-magnetic sphere with a radius of 85 mm in vacuum. The sphere's relative material parameters are $\varepsilon = \mu = 4$. Here, the electric field distributions are displayed corresponding to: (a) electric dipole (ED), (b) magnetic dipole (MD), (c) electric quadrupole (EQ), (d) magnetic quadrupole (MQ), (e) electric octupole (EO), (f) magnetic octupole (MO), (g) electric hexadecapole (EH), and (h) magnetic hexadecapole (MH). The sphere is illuminated by a monochromatic plane wave, linearly polarized along the $x$-axis and propagating along the $z$-axis, with a wavelength of 427 mm, corresponding to the quadrupole resonance. The power intensity of the incident wave is $I = 1 \, \mathrm{W/m^2}$.}
    \label{fig:FIG3}
\end{figure}

These types of analysis highlight the advantages of the spherical multipole decomposition technique developed in this work, which is based on expanding the electromagnetic fields outside the particle into vector spherical harmonics. Unlike other conventional approaches that rely on asymptotic approximations valid only in the far-field region — thus limiting the analysis to far-field quantities — our method provides a complete and exact multipole decomposition of the electromagnetic fields across all three spatial regions external to the scatterer: the reactive near-field region, the radiating near-field (Fresnel) region, and the far-field (Fraunhofer) region. In addition, our theoretical framework is not limited to isotropic spheres; it can be applied to anisotropic spheres and even to arbitrarily shaped scatterers composed of anisotropic materials.

\section*{Expressions for the Anisotropic Permeability of Ferrites}

For completeness, in this section we briefly discuss the anisotropic permeability tensor of a ferrite in the presence of a static magnetic field (DC) oriented along the $z$-axis, i.e., $ \hat{\mathbf{H}} = H_0 \hat{z} $, where $H_0$ denotes the strength of the applied field.\cite{pozar_2012} The ferrite interacts with an alternating current (AC) microwave propagating along the $z$-axis, whose magnetic field lies in the $xy$-plane, expressed as $ \mathbf{H} =  H_x \hat{\mathbf{x}} + H_y \hat{\mathbf{y}} $. Under the assumption that the magnetic field strength of the AC microwave is small compared to the DC magnetic field, the permeability tensor of the ferrite is given by:

\begin{equation}
\hat{\mu}_p = \mu_{0} \begin{pmatrix}
\mu^{T} & i \kappa & 0 \\
-i \kappa & \mu^{T} & 0 \\
0 & 0 & 1 \\
\end{pmatrix}.
\end{equation}

The elements of the permeability tensor are complex functions of the angular frequency $ \omega $ of the AC microwave. Consequently, they can be expressed in the standard form:

\begin{equation}
\mu^{T} \left( \omega \right) = 1 + \chi^{\prime} \left( \omega \right) + i \chi^{\prime \prime} \left( \omega \right), ~~~~~~~   \kappa \left( \omega \right) = \Omega^{\prime} \left( \omega \right) + i \Omega^{\prime \prime} \left( \omega \right).
\end{equation}

The real and imaginary parts of the diagonal element $ \mu^{T} \left( \omega \right) $ are given by:

\begin{eqnarray}
\chi^{\prime} \left( \omega \right) &=&  \frac{\omega_0 \omega_m \left[ \omega_0^{2} - \omega^{2} \left( 1 - b^2 \right) \right]}{D}, \\
\chi^{\prime \prime} \left( \omega \right) &=&  \frac{\omega \omega_m b \left[ \omega_0^{2} + \omega^{2} \left( 1 + b^2 \right) \right]}{D}.
\end{eqnarray}

Here, $ \omega_0  =  \mu_0 \gamma H_0 $ is the so-called Larmor frequency, $ \gamma = 1.759 \times 10^{11} $ C/kg is the gyromagnetic ratio, and $ \omega_m = \mu_0 \gamma M_0  $ is the magnetization frequency, where $ M_0 $ denotes the saturation magnetization. The parameter $ b = \mu_0 \gamma \Delta H / \left( 2 \omega \right)  $ represents a dimensionless damping constant, and $ \Delta H $ is the linewidth of the gyromagnetic resonance. The denominator $D$ is given by:

\begin{equation}
D = \left[ \omega^{2}_{0} - \omega^{2} \left( 1 + b^2 \right) \right]^2 + 4 \omega_{0}^{2} \omega^{2} b^{2}.
\end{equation}

Similarly, the real and imaginary parts of the off-diagonal elements $ \kappa \left( \omega \right) $ are given by:

\begin{eqnarray}
\Omega^{\prime} \left( \omega \right) &=& \frac{\omega \omega_m \left[ \omega_{0}^{2} - \omega^{2} \left( 1 + b^2 \right)  \right]}{D}, \\
\Omega^{\prime \prime} \left( \omega \right) &=& \frac{2 \omega^{2} \omega_m \omega_{0}}{D}.
\end{eqnarray}

Magnetization values reported in the literature are commonly expressed in Gaussian units. To convert them to the International System of Units (SI), the following transformation must be applied:

\begin{equation}
\hat{M}^{\text{SI}} =  \left( 10^{-4} \frac{\text{Wb}}{m^2} \right)  \frac{4 \pi}{\mu_0} \hat{M}^{G},
\end{equation}

where $ \hat{M}^{\text{SI}} $ and $ \hat{M}^{G} $ represent the magnetization in SI and Gaussian units, respectively. The conversion factor in parentheses corresponds to $ 1 \left[ G \right] = 10^{-4} \text{Wb}/m^2 $. Similarly, the magnetic field strength $ H $, measured in oersteds (Oe), must be converted using:

\begin{equation}
1 \left[ \text{Oe} \right] = \left( \frac{10^{3}}{4 \pi} \right) \left[ \frac{A}{m} \right].
\end{equation}

This ensures consistency with the SI unit system.

\section{Acknowledgment}
This work has been funded by the following research projects: Brasil 6G Project with support from RNP/MCTI (Grant 01245.010604/2020-14), xGMobile Project code XGM-AFCCT-2024-3-1-1 with resources from EMBRAPII/MCTI (Grant 052/2023 PPI IoT/Manufatura 4.0) and FAPEMIG (Grant PPE-00124-23), SEMEAR Project supported by FAPESP (Grant No. 22/09319-9), SAMURAI Project supported by FAPESP (Grant 20/05127-2), Ciência por Elas with resources from FAPEMIG (Grant APQ-04523-23), Fomento à Internacionalização das ICTMGs with resources from FAPEMIG (Grant APQ-05305-23), Programa de Apoio a Instalações Multiusuários with resources from FAPEMIG (Grant APQ-01558-24), and Redes Estruturantes, de Pesquisa Científica ou de Desenvolvimento Tecnológico with resources from FAPEMIG (Grant RED-00194-23). Authors wish also acknowledge the financial support from the Brazilian agencies National Council for Scientific and Technological Development-CNPq (308068/2025-4, 303282/2021-5, 405014/2025-2), FAPEMIG (APQ-02782-25), and FAPESP (2018/22214-6 and 2025/16174-5). N. M. acknowledges support from the European Research Council (ERC Starting Grant No. 101116253), the Swedish Research Council (Grant No. 2021-05784) and the Knut and Alice Wallenberg Foundation through the Wallenberg Academy Fellows Program (Grant No. 2023.0089).

\section{Disclosures}
The authors declare no conflicts of interest.


\bibliography{sample}

\end{document}